\documentclass[12pt, a4paper]{article}
\usepackage[english]{babel}
\usepackage{graphicx}
\usepackage{url}
\usepackage{framed}
\usepackage[font=small, labelfont=bf]{caption}
\usepackage[normalem]{ulem}
\usepackage{amsmath}
\usepackage{amsthm}
\usepackage{amssymb}
\usepackage{amsthm}
\usepackage{amsfonts}
\usepackage{enumerate}
\usepackage{bbold}
\usepackage{algorithm}
\usepackage[noend]{algpseudocode}
\usepackage[utf8]{inputenc}
\usepackage{dsfont}
\usepackage[toc]{appendix}
\usepackage{algorithm}
\usepackage[noend]{algpseudocode}
\usepackage{subfigure}
\usepackage[top=1 in,bottom=1in, left=1 in, right=1 in]{geometry}
\newcommand{\argmin}{\operatornamewithlimits{argmin}}
\algrenewcommand\algorithmicrequire{\textbf{Input:}}
\algrenewcommand\algorithmicensure{\textbf{Output:}}
\algnewcommand\algorithmicinput{\textbf{Input:}}
\algnewcommand\algorithmicoutput{\textbf{Output:}}
\algnewcommand\Input{\item[\algorithmicinput]}%
\algnewcommand\Output{\item[\algorithmicoutput]}%

\theoremstyle{definition}
\newtheorem{definition}{Definition}[section]

\newcommand{\distas}[1]{\mathbin{\overset{#1}{\kern\z@\sim}}}%

\newcommand{\R}{\mathbb{R}}

\newcommand{\Z}{\mathbf{Z}}
\newcommand{\X}{\mathbf{X}}
\newcommand{\SID}{\mathcal{X}_{SID}}
\newcommand{\datasetX}{\mathcal{X}}
\newcommand{\datasetY}{\mathcal{Y}}
\newcommand{\om}{\omega_{\mathcal{X}_{SID}}}

\newcommand*\ERR{\mathbf{\mathcal{E}}}

\theoremstyle{definition}
\newcommand{\condition}{{\ERR|{\X=x}}}

\title{Mape\_Maker: A Scenario Creator}
\date{\today}
\author{Guillaume Goujard --  University of California Berkeley\\
  Jean-Paul Watson -- Sandia National Laboratories\\
  David L.\ Woodruff -- University of California Davis}

\begin{document}

\maketitle
\begin{abstract}
We describe algorithms for creating probabilistic scenarios for the 
situation when the underlying forecast methodology is modeled as being 
more (or less) accurate than it has been historically. Such scenarios
can be used in studies that extend into the future and may need to 
consider the possibility that forecast technology will improve. Our
approach can also be used to generate alternative realizations of 
renewable energy production that are consistent with historical forecast 
accuracy, in effect serving as a method for creating 
families of realistic alternatives -- which are often critical in 
simulation-based analysis methodologies.
\end{abstract}

\section{Introduction}

Uncertainty associated with the forecasted output of renewable energy 
sources such as wind and solar mandates analysis and management techniques 
that take stochastics into account. A growing literature describes 
methods for creating and evaluating probabilistic {\em scenarios}, 
which are forecasts of renewables power generation with an attached 
probability. A representative sample of this literature can be found in
\cite{KirschenPES2014,pinsoneval,pinsonmadsen,PinsonGirard2012,sarietal,WOODRUFF2018153}.
Here, we are interested in creating probabilistic scenarios for the 
situation when the underlying forecast methodology is modeled as being 
more (or less) accurate than it has been historically. Such scenarios
can be used in studies that extend into the future and may need to 
consider the possibility that forecast technology will improve. Our
approach can also be used to generate alternative realizations of 
renewable energy production that are consistent with historical forecast 
accuracy, in effect serving as a method for creating 
families of realistic alternatives -- which are often critical in 
simulation-based analysis methodologies. A general open-source software 
implementation of the methods described here -- a package called 
{\em mape\_maker} -- is publicly available at \url{https://github.com/mape-maker/mape-maker}.

\begin{figure}[h]
    \centering
    \includegraphics[scale=0.35]{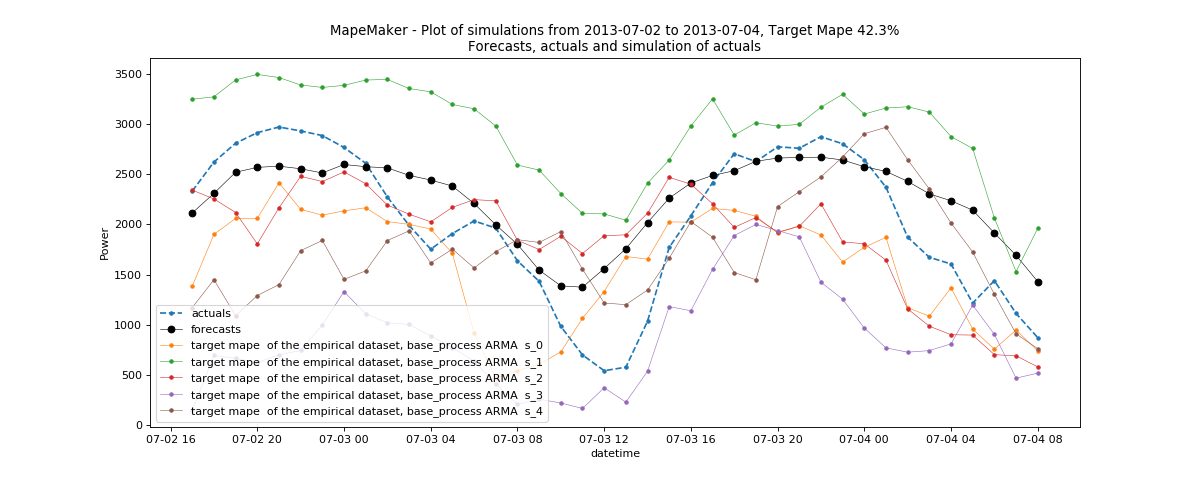}
    \caption{Illustration of 5 scenarios of wind production in CAISO representing alternative actuals. The forecast and realized actuals are also shown.}
    \label{fig:exampleofactuals}
\end{figure}

Given a time series of forecasts (e.g., daily over a year), we create a 
set of scenarios for renewable power production  that, based on a forecast
system with a specified accuracy, could reasonably correspond to the
forecasts. We often refer to these scenarios as {\em actuals}, to 
distinguish these values from historical forecasts. We can also create a 
set of forecasts that could reasonably correspond to a given time series 
of actuals. In other words, the process can be inverted. The correspondence 
between forecasts and actuals is based on analysis of historic forecast 
error distributions.  Subsequently, the word ``reasonably'' is replaced
with mathematical criteria concerning the error distribution, temporal 
correlation, and in the case of the forecast, curvature. As a preview of 
the output of this capability, consider Figure~\ref{fig:exampleofactuals}.
This figure provides a simple example where a set of 5 alternative 
``actual'' scenarios are constructed for a few days in July of 2013
based on wind forecast error data from obtained from the California 
Independent System Operator (CAISO) in the US for July 2013 through May
2015.  The target error -- specifically, the mean absolute percentage error
or MAPE -- is the value that was realized in the forecast error data. 
Because the scenarios are created for days in the past, we are able so 
show both the forecast and realized actuals on the same plot as the 
constructed scenarios.

\subsection{Measures of Forecast Error}

Let $(x_i)_i \in \R^n$ and $(y_i)_i \in \R^n$ denote two time-series. For
simplicity, we subsequently refer to these time-series as $x$
and $y$. We then define the following functions:
$$\begin{array}{cccccc}
RE & : & \mathbb{R}^*\times\mathbb{R} & \to & \mathbb{R} &\mbox{(Relative Error)} \\
 & & x, y & \mapsto & \frac{y - x}{x} \\
\end{array}$$
$$\begin{array}{cccccc}
MARE & : & {\mathbb{R}^*}^n\times\mathbb{R}^n & \to & {\mathbb{R}}_{+} &\mbox{(Mean Absolute Relative Error)} \\
 & & x, y & \mapsto & \sum_{i=1}^{n}\frac{|RE(x_i,y_i)|}{n} \\
\end{array}$$
The MAPE (Mean Absolute Percentage Error) is simply the MARE (Mean
Absolute Relative Error) given as a percentage. Our software library
communicates with users in terms of MAPE, but in our discussions here
it is convenient to use MARE and sometimes MAE (Mean Absolute Error)
variants.

While MAPE is a very popular way of characterizing forecast accuracy for 
renewables production, it is well-known to have a number of undesirable 
properties (see, e.g., \cite{badmape}).  One undesirable properties is 
that $x$ values of zero must be ignored in the calculation. We have 
organized our methods in such a way as to avoid division by zero. Most 
of the development here is based on converting the MAPE target to an
absolute error conditional on the value of $x$, so it would be relatively 
straightforward extension to convert our algorithms to use some measure 
of accuracy other than the MAPE.

\subsection{Notation Scheme}

We use $\datasetX$ and $\datasetY$ to denote paired input data of
length $n$. Note that which of these pairs is the forecast and which
is the actual depends on the user objective, i.e., what is being
simulated. For example, if one desires to obtain alternative actuals
from forecasts, then $\datasetY$ will be simulated actuals. Recalling
the canonical goal of constructing a vector $Y$ from $X$ input for a
specified range of dates, we use $\SID$ to denote the input data upon
which the construction is based (Simulation Input Data). It may or may
not be the case that $\SID$ is a subset of $\datasetX$. For the next
few sections, we assume that both input datasets are sorted according
to the $X$ values, e.g., $\datasetX = (x_i)_{i \leq n}$ with $n$ equal
to the cardinality of $\datasetX$ such that $\forall i < j, \: x_i <
x_j$. We will return to a temporal sorting
in Section~\ref{sec:Simulating} when we consider auto-correlation.
 
We use bold upper case font to denote random variables. As indicated
above, the role of the forecasts and the actuals can be reversed. If we
want to compute $y$, then $x$ is the input data for the simulation. We 
let $\ERR$ denote a random vector of errors such that $\ERR_i =
\mathbf{Y}_i-x_i$ so $$\mathbf{Y}_i = x_i + \ERR_i, \qquad \forall i \leq n.$$

We let $\varepsilon$ denote a vector of observed errors. We will focus on 
the modeling of $\mathbf{\varepsilon}$ in the following. The title of the
paper and the name of our software library derives from the requirement that 
simulated values $\tilde{y}$ must result in a MAPE close enough to the target
MAPE. We formalize this constraint as
$$\mathbb{E}[MARE(x,\mathbf{\tilde{Y}})] = \tilde{r},$$
where $\tilde{r}$ is the target MAPE divided by 100\%.

\subsection{Plausibility Criteria \label{sec:plauscrit} }
 
A main theme underlying this work that we will use to justify some of
our design choices involves what we refer to as {\em plausibility criteria}. 
For any requested MAPE, the distributions of errors computed should be 
as close as possible to the original error distributions while satisfying 
the target MAPE. If a user were to select the estimated MAPE as the requested
one, one would naturally expect the distribution of errors drawn from
the simulated distributions to be somehow ``close'' to the estimated distribution. 
For example, if the system of forecasts is producing a wide range of errors 
at very low forecasted power output, then even if the forecast technology is 
improving one would expect it to still produce a relatively wider range
of errors at low power regardless of the requested MAPE.  We formalize these criteria as 
follows in Definition~\ref{def:plausible}.

\begin{definition} \label{def:plausible}
A scenario set is said to be {\em plausible} if:
\begin{enumerate}
    \item The error distribution for the set is close to the empirical distribution of errors, i.e, its plausibility score is close to 1 (as defined in later in Section \ref{plausibity_score}).
    \item The computed auto-correlation coefficients for the set are close the empirical values, 
    \item the computed curvature for the set is close to the empirical value, especially when the scenarios are forecasts (because we observe that forecasts
      typically have lower curvature than actuals.)
\end{enumerate}
\end{definition}

\section{Modeling the Joint Distribution of $(\ERR, \X)$}

Let us define $\Z= (\ERR, \X)$. Here, $\Z$ denotes a random variable with
values in $(-\infty,+\infty) \times (0,+\infty)$ -- or, if the production 
capacity $cap$ is known by the forecaster, values in $[-cap,cap] \times
[0,cap]$. We denote by $f_{\Z}$ the density of $\Z$, and denote by $f_\ERR$ 
and $f_\X$ the marginals of $f_{\Z}$. Then,
  $$f_\ERR(\varepsilon) = \int_{-\infty}^{\infty}f_{\Z}(\varepsilon,x) dx, \qquad f_\X(x) = \int_{-\infty}^{\infty}f_{\Z}(\varepsilon,x) d\varepsilon$$
We also define the conditional density of $\ERR$ given $\X = x$ as:
  $$f_{\condition}(\varepsilon) = \frac{f(\varepsilon,x)}{f_\X(x)}$$

Modeling the conditional distribution of errors is important as these 
distributions can vary significantly with the value of input data. For
example, when the forecasts and the actuals are both low, the errors 
will be biased because the power cannot be below zero. Symmetrically, 
close to the maximum capacity, $cap$, errors are bounded by the fact 
that power cannot exceed maximum production capacity.

In this context, we introduce the functional $m(x)$ to denote the expected 
value of the absolute error of the distribution conditioned on $x$, defined as:
$$ m(x) = \mathbb{E}[|\ERR| \: | \X=x] = \int_{\varepsilon = -\infty}^{\infty}|\varepsilon|f_{\condition}(\varepsilon) d\varepsilon$$ 
 
We then introduce $r$ to denote the mean absolute relative error, defined as:
$$r = \mathbb{E}[\mathbb{E}\big[\frac{|\ERR| \: | \X}{\X}\big]] = \mathbb{E}[\frac{m(\X)}{\X}]$$

In Figure~\ref{fig:joint}, we provide an illustrative visualization of the 
relative error $RE$ as a function of actuals. We note that because actuals
are correlated with forecasts, the figure would be very similar if forecasts 
were used in instead. The data is for CAISO wind power data, ranging from 
July 1, 2013 to June 30, 2015. We will use this dataset for illustration 
throughout the paper, and refer to it informally as the {\em CAISO Wind} 
data set. These data are available in the mape\_maker software distribution;
the file is 
\verb|wind_total_forecast_actual_070113_063015.csv|.

\begin{center}
     \includegraphics[scale=0.5]{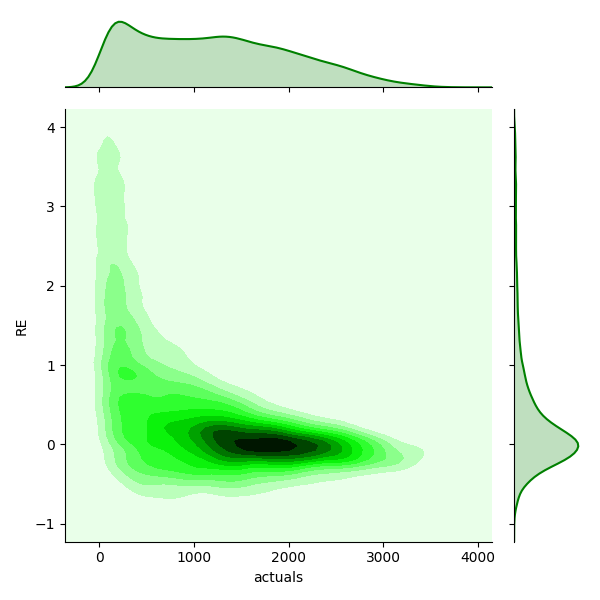}
    \captionof{figure}{Empirical joint distribution of $(\frac{\ERR}{\X}, \X)$ - CAISO Wind Power}
    \label{fig:joint}
\end{center}

\subsection{Estimating the Conditional Distribution of $\ERR|\X$, $\hat{f}_{\ERR|\X=x}$}

Given the notation $x \in \mathcal{X}$, we use the beta distribution on $[l,
  s+l]$ to model $f_{\ERR|X=x}$. In addition to the $l$ and $s$ that we will
refer to location parameters, a beta distribution requires two additional 
parameters -- $\alpha$ and $\beta$, i.e., the shape parameters. We chose the
beta distribution because it has finite support that helps us avoid power 
values below zero or above $cap$ and because the shape parameters provide the
flexibility necessary to model different behaviors for each $x$. We then
define 
$$\boxed{f_{\ERR|\X=x}(\varepsilon) = beta(\varepsilon; \: (\alpha, \beta, l, s)) = \frac{(\frac{\varepsilon-l}{s})^{\alpha - 1}(1- \frac{\varepsilon-l}{s})^{\beta - 1}}{B(\alpha, \beta)} }$$
with 
$$B(\alpha, \beta) = \int_{\varepsilon=l}^{l+s}(\frac{\varepsilon-l}{s})^{\alpha - 1}(1- \frac{\varepsilon-l}{s})^{\beta - 1} d\varepsilon$$

\subsection{Intervals for Conditional Estimation}

We now define a rule that will be used to estimate the parameters of the 
conditional density based on each $x$ of the input dataset. We choose to
take a fraction $a$ (e.g., 0.05) of data before and after each $x$. Let 
$G_X$ denote the empirical cumulative distribution function. Then, let 
$I_x^a = [G_X^{-1}(G_X(x)-a), \: G_X^{-1}(G_X(x)+a)]$. Thus $I_x^a$ is 
centered on $\bar{x}(x; \:a)=\frac{G_X^{-1}(G_X(x)-a)+G_X^{-1}(G_X(x)+a)}{2}$ 
with $2a$ fraction of the data. We fit the parameters on the sample 
$E_{I_x^a} = \{ \varepsilon_i, \; 1 \leq i \leq n,   x_i \in I_x^a,   \}$.
Note that for production values near zero and near the capacity, there 
could be as few as $a$ fraction of the values used.

To compute the estimation for a particular value $x'$, our method uses 
the interval $I_x^a$ for which $\bar{x}(x;a)$ is closest to $x'$ and 
uses the corresponding set $E_{I_x^a}$ to compute the parameters for
$x'$. For $x' \in \datasetX$ that are not close to zero or $cap$, the
closest $\bar{x}(x;a)$ to $x'$ will often be just $\bar{x}(x';a)$.
However, for very small or large values of $x'$ and when $\SID \not
\subset \datasetX$, the use of the interval with the closest mean is
most appropriate.

We will now describe how our method fits the parameters of the beta
distributions. Because every estimated quantity will depend on $a$,
we drop $a$ as a subscript or function parameter for notational
simplicity.

\subsection{Fixing $l$, $s$ and Estimating $\alpha$,  $\beta$ \label{sec:estimation}}

\subsubsection{Constraints on the Location Parameters \label{bounds}}

An informed choice of the location parameters will avoid simulating errors
leading to $y$ values lower than $0$ or greater then the $cap$ of the
dataset. We now define the function $y_{max}$, which returns the
maximum possible simulated value at $x$ according to a conditional
distribution $f_{\ERR|\X=x}$. Because the inverse of the corresponding
cumulative distribution function (CDF) evaluated at one, $F^{-1}_{\ERR|\X=x}(1) = l+s$, 
is the maximum of the error simulated; $F^{-1}_{\ERR|\X=x}(0) = l$ is
the minimum; 
and because we want to avoid simulating values above the cap or below zero we have
\begin{align*}
  y_{max}(x) &= x +F^{-1}_{\ERR|\X=x}(1) \\
  &= x+s+l \\
  &\leq cap.
\end{align*}

Similarly,
\begin{align*}
  y_{min}(x) &= x+F^{-1}_{\ERR|\X=x}(0) \\
  &= x+l\\
  &\geq 0.
\end{align*}

These two conditions give
\begin{align*}
  l &\geq -x \\
  s &\leq cap - x - l.
\end{align*}

Thus, we can define the estimators of the location parameters for each $x$ as:
\[\hat{l}(x) = \left\{ 
\begin{array}{l l}
    -x & \quad \text{if $min\big( {\varepsilon_i, \: x_i \in I_x} \big) \leq -x$} \\
    min\big( {\varepsilon_i, \: x_i \in I_x} \big) & \quad \text{else}\\
 \end{array} \right. \]
 
 \[ \hat{s}(x) = \left\{ 
\begin{array}{l l}
   cap - x - \hat{l}(x) & \quad \text{if $max\big( {\varepsilon_i, \: x_i \in I_x} \big) \geq cap-x$} \\
    max\big( {\varepsilon_i, \: x_i \in I_x} \big) - \hat{l}(x) & \quad \text{else}\\
 \end{array} \right. \]

\subsubsection{Choosing the Shape Parameters by the Method of Moments \label{moments}}
The mean and variance of a beta($\alpha$, $\beta$, $l$, $s$) distribution are:

\begin{align*}
  \mu &= \frac{s\alpha}{\beta+\alpha} + l\\
  V &= \frac{1}{s^2}\frac{\alpha\beta}{(\alpha+\beta)^2(\alpha+\beta+1)}\\
\end{align*}

We can now choose shape parameters by solving these two equations for $\alpha$ and $\beta$
\begin{align*}
  \hat{\mu}(x) &= \frac{\hat{s}(x)\alpha}{\beta+\alpha} + \hat{l}(s)\\
  \hat{V}(x) &= \frac{1}{\hat{s}(x)^2}\frac{\alpha\beta}{(\alpha+\beta)^2(\alpha+\beta+1)}\\
\end{align*}
to obtain $\hat{\alpha}(x)$ and $\hat{\beta}(x)$.

For any $x \in \datasetX \cup \SID$ assign 
$$ \hat{\mathcal{S}}_x = (\hat{\alpha}(x), \hat{\beta}(x), \hat{l}(x), \hat{s}(x)) $$

\subsection{Selecting $a$ \label{a_selection}}

We now develop an empirical way to select the best $a$. If $a$ is
small, the sample on which to fit the distribution will be small since
$I_x^a$ is small. Fitting a distribution on very little data is of
course dangerous. On the other hand, if $a$ is large, then the sample 
is too large to provide us with an estimation of the conditional density.
In the extreme where $a=1$, every conditional density will be equal to the
density of the relative error.

One way to select $a$ is to compute a discrepancy score between the
empirical distribution function and the one obtained by
estimating each conditional distributions with $2a$ of the data. Let $g$
be the empirical joint density of $(X,\varepsilon)$. Let $\hat{f}$ be
the joint density of $(\X,\ERR)$ taken as $\hat{f}_a(x,\varepsilon) =
\hat{f}_{\X}(x)*\hat{f}^a_{\ERR|\X=x}(\varepsilon)$.  We choose $a$ to
minimize the deviation between the real density and the simulated
density:
$$ D^2(a) = \int_x\int_{\varepsilon}(g(x,\varepsilon) - \hat{f}_a(x,\varepsilon))^2 d\varepsilon dx. $$

\section{Adjusting the conditional densities to fit a MARE target}

We will use a tilde to specify the distributions and variables that we
are simulating. 

\begin{itemize}
\item While $\ERR$ is the random variable of the error with
  properties that can be estimated from $\datasetX$, $\tilde{\ERR}$ is
  the random variable of error defined by a distribution that
  we will develop with desired properties for the simulation.
\item We make use of three conditional distributions : the population
  density,  $f_{\condition}$, the estimated density
  $\hat{f}_{\condition}$, and a simulation density,
  $\tilde{f}_{\condition}$.
\end{itemize}

We are now interested in modeling the conditional
distribution of $\tilde{\ERR}|\X$ so that the expected relative
absolute error of the simulated random variable $\tilde{\ERR}$ is : 

$$\mathbb{E}_{\tilde{\ERR}}[\frac{1}{n_{SID}} \sum_{x\in\SID}\frac{|\tilde{\ERR}| \: | \X=x}{x}\big] = \tilde{r} $$

\subsection{Adjusting the shape parameters so that it fits a target MAE \label{adjust}}

We want to adjust each conditional distribution so that the global
distribution of $\tilde{\ERR}$ satisfies the targeted MARE and so that they
keep the same shape parameters as the original distributions. To do
this we compute analytically the mean absolute error of a beta distribution
when $\alpha$ and $\beta$ are fixed. Let $l<0$ and $s+l>0$. Let
$b(\cdot; \alpha, \beta, l,s)$ be an arbitrary beta density function with parameters $(\alpha,
\beta, l, s)$ for which we define a  mean absolute error function of $l$ and $s$ given
values for $\alpha$ and $\beta$ as
\begin{align*}
  \nu(l,s; \alpha, \beta)  &= \int_{\varepsilon = l}^{s+l}|\varepsilon|b(\varepsilon;  \alpha, \beta, l, s) d\varepsilon. \\
\end{align*}

 We will make two remarks:
 $$\lim_{s\to0}  \nu(l,s; \alpha, \beta) = 0, \quad \forall l < 0 $$ 
 $$ \nu(l,s; \alpha, \beta) \underset{s \xrightarrow{}\infty}{\sim} \frac{s\alpha}{\alpha+\beta} $$
  
Since $\nu$ is continuous ( it is a sum of continuous functions ), the
intermediate value theorem applies which means that $\nu(l,s; \alpha, \beta)$ can
achieve any value and in particular, the value needed to in order to hit
the specified error target. 
 
 Thus, once we are given $\alpha$, $\beta$, and a target value for the
 absolute error at a particular value of $x$, we need to find the
 intersection between a hyperplane defined by the target and the
 surface defined by $\nu(l,s; \alpha, \beta)$ to establish values for
 $\tilde{l}$ and $\tilde{s}$.  For $x \in
 \mathbb{R}_+$ we will want to choose the solution that minimize the
 distance to the estimated values $\hat{l}(x) $ and $\hat{s}(x)$ while
 hitting a target mean absolute error $m(x)$ and without
 changing the shape parameters. 
 
 \begin{equation} \label{eq:w_prog}
\begin{aligned}
(\tilde{l}(x), \tilde{s}(x)) =& \argmin_{l,s} & & (l -\hat{l}(x) )^2 + (s -\hat{s}(x))^2 \\
& \text{s.t.} & &  l \in \R, \: s \in \R_+ \\
& & & 0 \geq l \geq - x \\ 
& & & 0 \leq s \leq cap - x - l \\
& & &  \nu(l,s; \hat{\alpha}(x), \hat{\beta}(x)) = m(x)
\end{aligned}
\end{equation}

However, in our case, since there are bound constraints on l and s
(see section~\ref{bounds}), $ \nu$ cannot hit every target $m(x)$. We
compute a maximum target function that can be hit as: $$m_{max}(x) =
\max_{l \in (-x,0],s \in [0, cap-x) } \nu(l,s; \hat{\alpha}(x),
\hat{\beta}(x))$$ The target function $m$ must then be bounded for
every $x$ by :
\begin{equation} \label{constraint_target}
\begin{aligned}
m(x) \leq m_{max}(x) 
\end{aligned}
\end{equation}
Given a mean absolute error target function $m$ satisfying inequality~(\ref{constraint_target}) we obtain for any $x$, a beta distribution of parameters $\tilde{\mathcal{S}}_{x, m} = (\hat{\alpha}(x), \hat{\beta}(x), \tilde{l}(x), \tilde{s}(x))$ that satisfies the mean absolute error target and that is the closest possible to the estimated distribution. We now proceed to allocate an error target to each $x \in \SID$ that we will call $\tilde{m}$ that depends on the target MARE and on a weight function.

\subsection{Changing the conditional distributions \label{sec:adjust}}

\subsubsection{Weight functions}

Let's define $\Omega_{\SID}$ as the set of functions $\om$ defined on
$\SID$ such that $$\frac{1}{n_{SID}} \sum_{x\in\SID}\om(x) =
1.$$ We call them {\em weight functions}.  Weight functions will be
used to assign a target MAE to obtain from each of the conditional
distributions $\tilde{\ERR}|\X=x$, for all $x \in \SID$. It can also
be seen as the function that weights the contribution of the Absolute
Error of each conditional distribution to the Mean Absolute Relative
Error of the simulation.

\subsubsection{Target function generator}

We also define the following functional that we call target function generator. 
$$\begin{array}{cccccc}
\tilde{m} & : & \SID \times \mathbb{R}_+ \times \Omega_{\SID} & \to & \mathbb{R}_+ &\mbox{(Target function generator)} \\
 & & x , \tilde{r}, \omega & \mapsto & \tilde{r}x\omega(x), \: x > 0  \\
 \end{array}$$
 
 For a fixed $\tilde{r}$ and $\omega$, $\tilde{m}(. \:, \: \tilde{r}, \omega)$ is a target function.
 Since the target function will be used to directly adjust the conditional distribution, it must respect the inequality~(\ref{constraint_target}). Finally, we say that a target mare $\tilde{r}$ is \textit{feasible} for a given $\omega \in \Omega_{\SID}$ if $$\forall x \in \datasetX, \: \tilde{m}(x, \tilde{r}, \omega) \leq m_{max}(x)$$

 \subsubsection{Zero power input} We recall that the zero input does not count in the computation of the MARE. However, we want the distribution of the simulated errors to be drawn from the estimated distribution. In other words :
$$\forall \tilde{r} \in \mathbb{R}_+, \quad \tilde{l}(0) = \hat{l}(0) \textit{ and } \tilde{s}(0) = \hat{s}(0) $$
We assign 
$$\tilde{m}(0) = \hat{m}(0) $$ To avoid big discontinuities in the
parameters of the beta distributions, we could take as $ \tilde{l}(0)
= \lim_{x \rightarrow 0}\tilde{\ell}(x)$, $ \tilde{s}(0) = \lim_{x
  \rightarrow 0}\tilde{s}(x)$,

\subsubsection{Convergence to the requested MARE \label{sec:convergence}}
 
Using the function $\tilde{m}$ to assign target MAE for each SID input
will allow us to hit the targeted MARE using the simulation
distribution. Indeed, let us define the random variable
$\tilde{\ERR}|\X$ with density $\tilde{f}_ {\condition}(\varepsilon) = b(\varepsilon,
\tilde{\mathcal{S}}_{x, \tilde{m}}), \:  \varepsilon \in
(-cap, cap)$. If we establish the distribution parameters as described in Section~\ref{adjust} and solve program~(\ref{eq:w_prog}) with $m(x) =  \tilde{m}(x ; \: \tilde{r}, \om)$ we have,
 \begin{equation*}
\int_{\varepsilon = -\infty}^{\infty}|\varepsilon|b(\varepsilon,
\tilde{\mathcal{S}}_{x, \tilde{m}})d\varepsilon = \tilde{m}(x ; \: \tilde{r}, \om), \quad \forall x \in \SID.
\end{equation*}
 Then, the expected MARE with the errors drawn from these distributions and with the inputs in the $\SID$ is :
\begin{align*}
	\mathbb{E}_{\tilde{\ERR}}[\frac{1}{n_{SID}} \sum_{x\in\SID}\frac{|\tilde{\ERR}| \: | \X=x}{x}\big] &= \frac{1}{n_{SID}} \sum_{x\in\SID} \frac{\mathbb{E}_{\tilde{\ERR}}[|\tilde{\ERR}| \: | \X=x]}{x}\\
	 &= \frac{1}{n_{SID}} \sum_{x\in\SID} \frac{ \tilde{m}(x ; \: \tilde{r}, \om)}{x}\\
	&= \frac{\tilde{r}}{n_{SID}} \sum_{x\in\SID} \om(x)\\
	&= \tilde{r} \\
\end{align*}

This is true with any weight function for which
$\frac{1}{n_{SID}} \sum_{x\in\SID}\om(x)  = 1$. We now proceed to describe the
construction of a sensible weight function.

\subsection{Weight function for $\SID = \datasetX$ \label{sec:weightfunction}}

We recall the plausibility criteria: we want our simulations of errors
to be as close as possible to the population distribution. In
particular, suppose that we want to do a simulation with a target MARE
that happens to be the same as the MARE for the original data
($\datasetX$ and $\datasetY$) and further suppose that we want to
simulate using values from the entire data set (i.e., $\SID =
\datasetX$). Then we expect the simulated conditional distributions to
be equal to the estimated conditional distributions. In other words,
$$\SID = \datasetX, \quad \tilde{r} = r_{\hat{m}} \implies \forall x
\in \datasetX, \tilde{l}(x) = \hat{l}(x) \textit{ and } \tilde{s}(x) =
\hat{s}(x) $$ Solving the linear program (\ref{eq:w_prog}) defined in
subsection \ref{adjust}, leads to $\tilde{l}(x) = \hat{l}(x) \mbox{
  and } \tilde{s}(x) = \hat{s}(x), \forall x \in \datasetX$.

If we define the following $\hat{\omega}_{\datasetX}$ function,
$$\forall x \in \datasetX, \: \hat{\omega}_{\datasetX}(x) := \frac{\hat{m}(x)}{xr_{\hat{m}}} = \frac{\int_{\varepsilon = -\infty}^{\infty}|\varepsilon|\hat{f}_{\condition}(\varepsilon) d\varepsilon}{xr_{\hat{m}}} $$

First, we can verify that we have $\frac{1}{n} \sum_{x\in\datasetX}\hat{\omega}_{\datasetX}(x)  = 1$. It is thus a weight function. 

The choice of this weight function is natural when $\SID = \datasetX$
because it is the ratio of the expected relative error simulated at
$x$ over the mean relative error when the errors are distributed
according to the estimated joint distribution. However, choosing it when $\SID
\not = \datasetX $ would satisfy our requirement for plausibility but
it would prevent us from hitting the requested MARE.

Figure~\ref{fig:CAISOratio} illustrates that for the full CAISO wind dataset, the weight function presents a hyperbolic shape.  The low values account for the biggest part of the MAPE.

\begin{center}
     \includegraphics[scale=0.5]{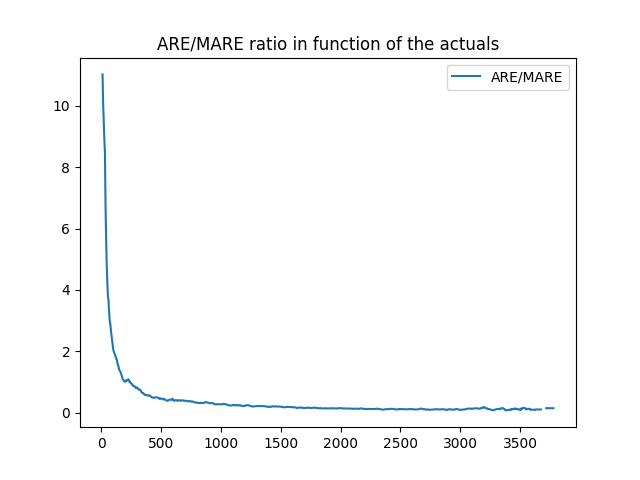}
     \captionof{figure}{$\hat{\omega}_{\datasetX}(x) =  \frac{\hat{m}(x)}{x\hat{r}} $ ratio for the CAISO wind
     dataset.}
    \label{fig:CAISOratio}
\end{center}

\subsection{Weight function for $\SID$ and an arbitrary $\tilde{r}$ \label{plausibity_score} }

Let us define the following Real that we call the plausibility score : 
$$P_{\SID} = \frac{1}{n_{SID}}
\sum_{x\in\SID}\hat{\omega}_{\datasetX}(x) $$ When $\SID \not=
\datasetX$, the distribution of the SID is different from the
distribution of the input dataset. Thus we do not necessarily have
$P_{\SID} = 1$. A goal of our method is to meet the requested MARE, at
least in expectation, no matter the $\SID$. If $P_{\SID}$ is greater
than 1, it means that the distribution of $\SID$ has more data in the
range where the weight function takes high values. This means that if
use $\hat{\omega}_{datasetX}$, we are going to simulate too many
errors with high values. While it has some physical sense, we are
nonetheless going to simulate a greater MAPE than
expected. Symmetrically, if $P_{\SID}$ is smaller than one, we are
going to retrieve a lower MAPE than expected. This is illustrated in
Figure~\ref{fig:compare_density}), the density for the $\SID$ between
December 2013 and March 2014 indicates more values at lower power than
for the entire dataset, $\datasetX$.  If we simply used
$\hat{\omega}_{\datasetX}(x)$, for $x \in \SID$, then meeting the
target AREs for each $x$ would result in a MARE much greater than
specified. In other words, since the ARE/MARE ratio is very high for
the low power input, and since these inputs are over represented under
the distribution of the December 2013 - March 2014 SID, we are going
to simulate too many errors with a high target of mean absolute error.
To meet the target MARE, a re-scaled weight function must thus be
computed.

Let us define the following SID weight function :

$$\forall x \in \SID, \: \tilde{\omega}_{\SID}(x) := \frac{\hat{\omega}_{\mathcal{X}}}{P_{\SID}}$$
With the re-scaled factor, we have $\frac{1}{n_{SID}} \sum_{x\in\SID}\tilde{\omega}_{\SID}(x) =1$ so $\tilde{\omega}_{\SID} \in \Omega_{\SID}$. 

Finally, for a given feasible $\tilde{r} \in \mathbb{R}_+$, we compute a $ \tilde{\omega}_{\SID}$ which allocates the absolute errors across $\SID$ based on the allocation from $\datasetX$. With these two parameters we can compute $\tilde{m}(x; \: \tilde{r}, \tilde{\omega}_{\SID}), \: x \in \SID$. According to Section~\ref{sec:convergence}, defining $\tilde{\ERR}$ from this target function, will get us $\mathbb{E}_{\tilde{\ERR}}[\frac{1}{n_{SID}} \sum_{x\in\SID}\frac{|\tilde{\ERR}| \: | \X=x}{x}\big] = \tilde{r} $.

We can also get the feasibility region for the target mare. For a given $\tilde{r}$ to be a feasible target mare, it must satisfy $\forall x \in \datasetX, \: \tilde{m}(x, \tilde{r}, \tilde{\omega}_{\SID}) \leq m_{max}(x)$. Thus the feasibility region is :

$$\tilde{\mathcal{R}}_{\SID} = P_{\SID} \min\big(\frac{m_{max}(s)}{s\hat{\omega}_{\mathcal{X}}(s)}, s \in \SID \big) $$

\begin{center}
     \includegraphics[scale=0.5]{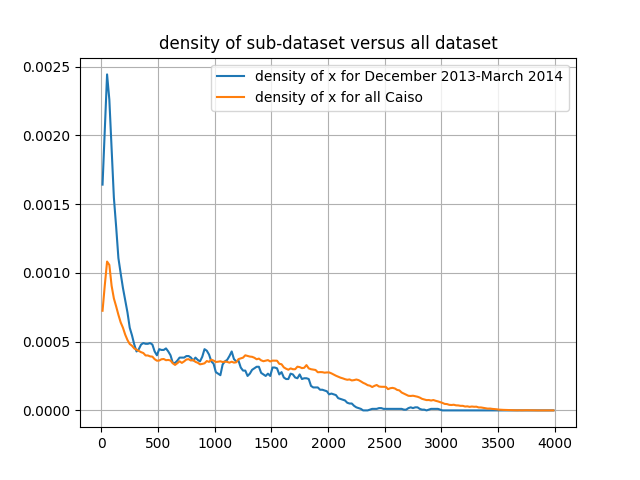}
    \captionof{figure}{Comparison test density versus all dataset density}
    \label{fig:compare_density}
\end{center}

\subsection{Simulating Without Auto-correlation \label{sec:Simulating}}

 This is now straightforward. First, to obtain a simulation of errors, we are simulating a vector identically and independently distributed uniformly on $[0,1]$, $(\tilde{\mathbf{U}}_t)_{t\leq n_{SID}}$. Then 
$$ \tilde{\mathbf{\ERR}}_t =  \tilde{F}_{\ERR|\X=x_t}^{-1}(\tilde{\mathbf{U}}_t), \quad \forall t \leq n_{SID}.$$
Let,
$$ \tilde{\mathbf{Y}}_t =  x_t + \tilde{F}_{\ERR|\X=x_t}^{-1}(\tilde{\mathbf{U}}_t), \quad \forall t \leq n_{SID}$$
so
$$\mathbb{E}[MARE(x,\mathbf{\tilde{Y}})] = \tilde{r}.$$

While we are hitting the target mare, the entire auto-correlation of
the errors simulated relies solely on the auto-correlation of the
input. In the extreme case where the errors are not depending on the
input i.e $\tilde{F}_{\ERR|\X=x_i} = \tilde{F}_{\ERR|\X=x_0}, \;
\forall i \leq n_{SID} $ - which is the case for the middle power
range for the CAISO wind data - then our simulations would have a null
auto-correlation function. Implementing a base process to replace
$(\tilde{\mathbf{U}}_t)_{t\leq n_{SID}}$ will generate the needed
auto-correlation to satisfy the second point of the plausibility
criteria.

\section{Inferring a Base Process \label{sec:Correlation}}

The idea is to simulate a Base Process  $\tilde{\mathbf{U}}_t$ of marginal Uniform in [0,1] depending on the past p lags $\tilde{U}_{t-i}, \; i \leq p$ and the past q lags of errors over the base process $\delta_{t-i}, \; i \leq q$ . Then, as previously in section~\ref{sec:Simulating}, we would simulate the errors via the transformation $\hat{F}_{\ERR|\X=x_t}^{-1}(\tilde{\mathbf{U}}_t)$.

We model $Z_t = \phi^{-1}(U_t) \in (-\infty, \infty)$ as a
Gaussian Process and more specifically as an ARMA process.
Heuristically, we will show that this method gets us a good
auto-correlation function for the simulations.

Inspired by the ARTA fit method (see \cite{biller2005}). We denote the
CDF for the standard normal distribution $\phi$ and the CDF of the
conditional distribution $\ERR|\X=x_t$, which is a beta distribution
fit using $\datasetX$, $\hat{F}_{\ERR|\X=x_t}$. Let us define the
following time-series $(\hat{Z}_t)_t$:

$$\forall t \leq n, \quad \hat{Z}_t = \phi^{-1}(\hat{F}_{\ERR|\X=x_t}(\varepsilon_t)) $$

We use the notation $(\hat{Z}_t)_t$ for the base process time-series
of the dataset. Its empirical distribution is close to a standard
Gaussian. Indeed, in section~\ref{sec:estimation} we are estimating
the conditional distribution so that $\ERR_t | \X=x_t$
has distribution that is approximated by $\hat{f}_{\ERR|\X=x_t}$, thus,
$\hat{F}_{\ERR|\X=x_t}(\ERR_t) \dot\sim \mathcal{U}[0,1]$ and
$\hat{\mathbf{Z}}_t \dot\sim \mathcal{N}(0,1)$.  We fit on this base
process an ARMA process. The standard definition of an ARMA process of
order $p$ and $q$ uses $(a_i)_{i \leq p}$ and $(b_i)_{i \leq q}$ as
coefficients so we temporarily reuse those symbols in this section.
    
    \begin{definition}
    $\{\Z_t\}$ is a base process if 
    \begin{itemize}
        \item $\{\Z_t\}$ follows an ARMA process of order $p$ and $q$
          : $$ \mathbf{Z}_t = \sum_{h=1}^p a_h \mathbf{Z}_{t-h} +
          \sum_{h=1}^q b_h \mathbf{\delta}_{t-h} +
          \mathbf{\delta}_t $$ Where $\{\mathbf{\delta}_t\}$ are the
          iid Gaussian error of mean 0 and variance $\sigma_\delta^2$.
        \item $Var[\mathbf{Z}_t] = 1$, $E[\mathbf{Z}_t]=0$, so that
          for all $t$, $\quad \mathbf{Z}_t \sim N(0,1)$.
    \end{itemize}
    \end{definition}
    
We run a grid search over multiple $(p,q)$ and we select the ARMA
model to minimize the BIC criterion. The $(a_i)_{i \leq p}$ and
$(b_i)_{i \leq q}$ found during the process define a function that
enables us to generate base processes which will create the
auto-correlation that we are looking for;  however,
fitting an ARMA imposes no constraint on the variance of noise
$\sigma_{\delta}$. So we are free to specify $\sigma_{\delta}$ so that we get
$Var[\tilde{\mathbf{Z}}_t] = 1$.

Then,  we can simulate directly the
error by 
$$ \tilde{\mathbf{\ERR}}_t =
\tilde{F}_{\ERR|\X=x_t}^{-1}(\phi(\tilde{\mathbf{Z}}_t)), \quad
\forall t \leq n_{SID}$$ and also get the result for
the expected MARE established in section~\ref{sec:Simulating}.

\section{Enforcing Curvature \label{sec:Curvature}}

Let $(y_i)_i \in \R^n$ denote an simulation output time series. We define
{\em curvature} at a point $i$ in $(y_i)_i$ as
$$s_i = y_{i+2}-2y_{i+1}+y_{i} \quad \forall i < n-2, \qquad$$
\noindent i.e., a second difference.

Methods described in Section~\ref{sec:Correlation} successfully model
temporal correlation between the errors while satisfying a target MARE. 
However, some scenarios might not ``look right'' because of their lack of
smooth curvature. This is especially unsatisfying in the case of forecasted
renewables power production, which are much less sharp and erratic when 
compared to actual quantities.

\begin{figure}[htp]
  \centering
  \subfigure[Illustration of forecast scenarios without curvature adjustment.\label{fig:nosmooth}]{\includegraphics[scale=0.4]{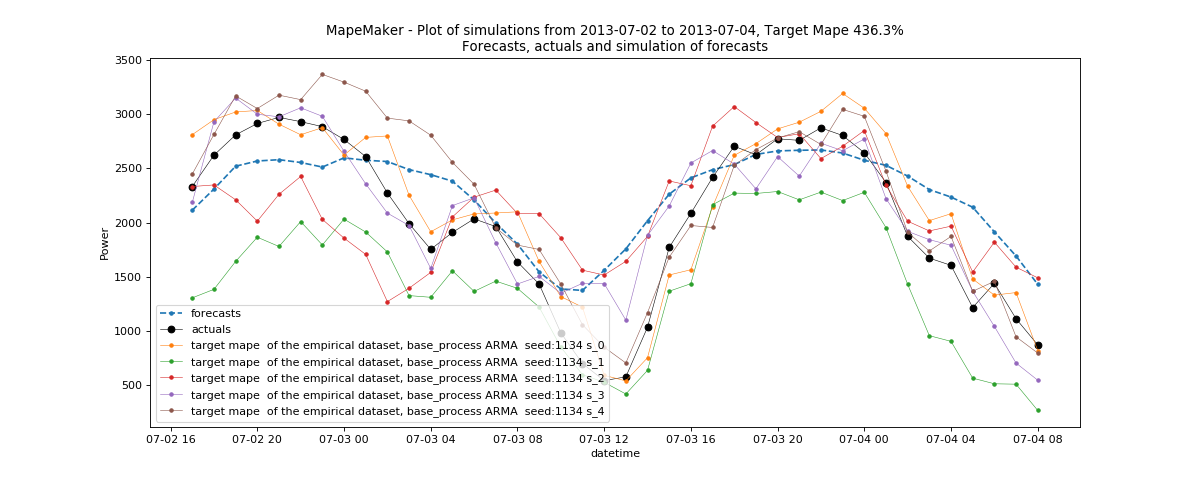}}
  \subfigure[Illustration of the same forecast scenarios with curvature adjustment \label{fig:withsmooth}]{\includegraphics[scale=0.4]{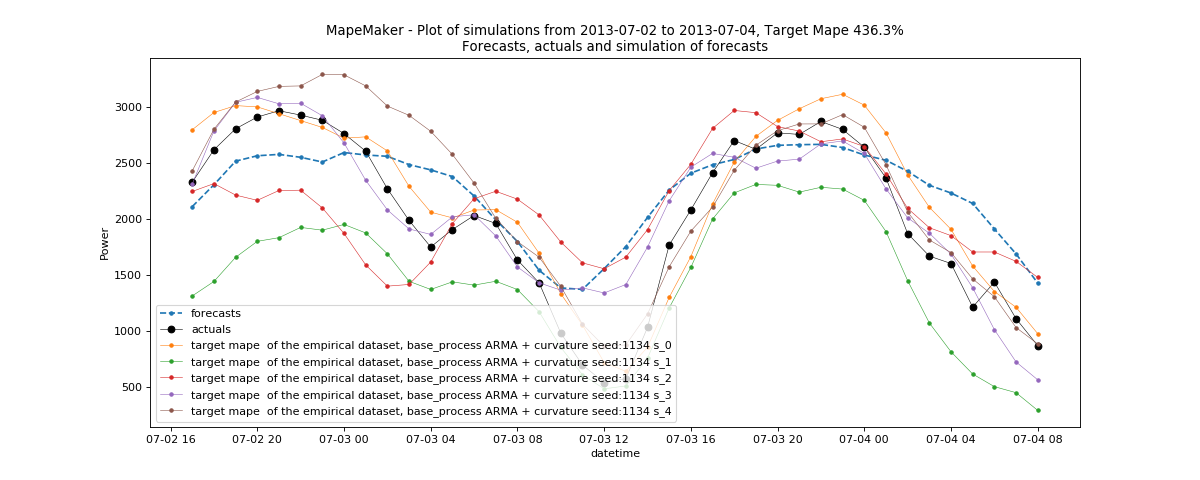}}\quad
\end{figure}

We now concretely illustrate this issue with analysis of the CAISO 
wind power production data introduced previously. In 
Figure~\ref{fig:nosmooth}, we show baseline scenarios resulting from
our proposed methods, i.e., without adjustment for curvature. As is
clearly observed, the simulated forecasts in this case closely mirror
the actuals -- and not the one ``true'' forecast. In contrast, we 
show in Figure~\ref{fig:withsmooth} a closely related set simulated
forecasts -- obtained by the procedure we now describe -- that instead 
exhibit significantly more smooth and realistic curvature. Ultimately,
the need for such adjustment depends entirely on the application.

In order to adjust the curvature of a forecast while still acheiving
a target MARE, one approach is to {\em a posteriori} adjust a time 
series that already satisfies a target MARE such that specific 
curvature characteristics are imposed. We now formalize this general
approach. 

We introduce a minimization problem in which we penalize deviations from
both a target second difference and the simulated forecast error. Per
earlier analysis, we can simulate $(\tilde{\varepsilon}_i)_i$ using an 
ARMA base process. Then, define $d \in \R_+$, and let $W_s$ and 
$W_\varepsilon$ denote user inputs in $\R_+$.

We then let $(y_i)_i$ denote the solution of the following 
mathematical program:
\begin{equation}
\begin{aligned}
& \underset{y}{\text{min}} & & \sum_{i=3}^n W_{s2}\bigg(|y_{i}-2y_{i-1}+y_{i-2}| - d \bigg)^2 + W_\varepsilon\bigg( y_i-x_i - \tilde{\varepsilon}_i \bigg)^2 \\
& \text{s.t.} & &  y \in [0,cap]^n \\
\end{aligned}
\end{equation}

For practical computation, we now transform this mathematical program so 
that the objective function is quadratic and constraints are linear -- such
that widely available mathematical programming solvers can be leveraged. 
The transformation yields the following equivalent mixed-integer linear
program (MILP), with $3n$ additional variables, $n$ equality constraints, 
and $3n$ inequality constraints ($6n$ if we consider that the three real 
vectors are negatively bounded by 0):
\begin{equation}
\label{eq:prog_curvature}
\begin{aligned}
& \underset{y, \: \lambda^+, \: \lambda^-, \: b}{\text{min}} & & \sum_{i=2}^n W_s\bigg(\lambda_i^+ + \lambda_i^- - d \bigg)^2 + W_\varepsilon\bigg(y_i-x_i - \varepsilon_i \bigg)^2 \\
& \text{s.t.} & &  y \in \R_+^n, \: \lambda^+ \in \R_+^n, \: \lambda^- \in \R_+^n , \: b \in \{0,1\}^n \\
& & & y_i \leq cap \\
& & & \lambda_i^+ - \lambda_i^-   =y_{i}-2y_{i-1}+y_{i-2}, \quad \forall i \leq n \\
& & & \lambda_i^+ \leq b_i d_{max} \\
& & & \lambda_i^- \leq (1-b_i) d_{max} \\
\end{aligned}
\end{equation}
where $d_{max}$ denotes a large constant; a safe value is $4cap$.

To verify equivalence of the two mathematical programs, we note that 
if $y_{i}-2y_{i-1}+y_{i-2} \geq 0$, then because $b_i \in \{0,1\}$, 
$\lambda_i^-$ is equal to 0 with the two last equations. Then 
$\lambda_i^+ =y_{i}-2y_{i-1}+y_{i-2}$ and $\lambda_i^+ +
\lambda_i^- = y_{i}-2y_{i-1}+y_{i-2}$. We use the same reasoning when
$y_{i}-2y_{i-1}+y_{i-2} < 0$.

Numerous open source and commercial solvers are available for such a
mathematical program. However, solution time does generally increase
with $n$. In many applications the restrictions on curvature are 
motivated by aesthetic or heuristic considerations. Thus, it can be
reasonable to specify a ``loose'' optimality gap to avoid excessive 
computation time.

\section{Putting it all Together}

In this section we summarize the process to deliver a simulation with correct targets.

\subsection{Procedures for estimation}

First, as shown in Algorithm~\ref{alg:euclid}, we preprocess the data
and estimate the conditional distributions using the methods explained
in section~\ref{sec:estimation}. This results in a set of beta
distribution parameters for each input from the whole dataset called
$\hat{\mathcal{S}}$. To estimate the parameters we recall that the
user should specify a data fraction (e.g., 0.05), for the sampling.
(The software provides an option to produce a curve for the scores
described in Section~\ref{a_selection}.)

\begin{algorithm}
\caption{Estimating the beta distributions}\label{alg:euclid}
\begin{algorithmic}[1]
\Input x, y, a\Comment{Input time-series and percent of data}
\Output $\hat{\mathcal{S}}_x$
\Procedure{Computing\_Estimation\_Parameters}{$x$, $y$, $a$}
\State $\datasetX\gets sort(x)$ 
\For{$x \in \datasetX$}\Comment{Applying the methodology explained in section~\ref{sec:estimation}}
\State Compute the interval of estimation $I^a_x$ and sample $E^a_x$ .
\State $\bar{x}(x,a)\gets \bar{I^a_x}$
\State $\hat{l}(\bar{x}(x,a)), \hat{s}(\bar{x}(x,a))\gets Bounds(E^a_x, x)$\Comment{See section~\ref{bounds}}
\State $\hat{\alpha}(\bar{x}(x,a)), \hat{\beta}(\bar{x}(x,a))\gets Moments(mean(E^a_x), std(E^a_x))$\Comment{See section~\ref{moments}}
\EndFor

\For{$x \in \datasetX$}\Comment{Take the closest computed point of estimation}
\State $x'\gets argmin |\bar{x}(x',a) - x |$
\State $\hat{\mathcal{S}}_x\gets (\hat{\alpha}, \hat{\beta}, \hat{l}, \hat{s})(\bar{x}(x',a) )$
\EndFor

\State \textbf{return} $\hat{\mathcal{S}}$
\EndProcedure
\end{algorithmic}
\end{algorithm}

Next, as shown in Algorithm~\ref{alg:estwgt}, we estimate the
partitioning of the mean absolute percent errors according to the
input and we encode this information in the weight function. An
important feature of this procedure is the computation of
$r_{\hat{m}}$ which is the expected mean absolute relative error from
the conditional distributions (which may be close in value to, but is
different from, $\hat{r}$.)  This procedure is explained in
section~\ref{sec:weightfunction}.

\begin{algorithm}
\caption{Estimating the weight function \label{alg:estwgt}}
\begin{algorithmic}[1]
\Input $\hat{\mathcal{S}}_x$, $\datasetX$
\Output $\omega_{\datasetX}$
\Procedure{Computing\_Estimated\_Weight\_Function}{$\hat{\mathcal{S}}_x$, $\datasetX$}
\State $r_{\hat{m}}\gets 0$
\For{$x \in \datasetX$}\Comment{Applying the methodology explained in section~\ref{sec:weightfunction}}
\State $ \hat{m}(x)\gets \int_{\varepsilon = -\infty}^{\infty}|\varepsilon|beta(\varepsilon; \mathcal{S}_{\mathcal{X}}(x) ) d\varepsilon$ .
\State $ m_{max}(x)\gets \max \nu(l,s, \hat{\alpha}(x), \hat{\beta}(x)) $\Comment{See constraints on target function~(\ref{constraint_target})}
\State $\omega_{\datasetX}(x)\gets \frac{\hat{m}(x)}{x}$
\State $r_{\hat{m}}\gets r_{\hat{m}} +\hat{m}(x) $
\EndFor
\State $r_{\hat{m}}\gets \frac{r_{\hat{m}}}{|\datasetX|} $
\State $\omega_{\datasetX}\gets \frac{\omega_{\datasetX}}{r_{\hat{m}}}$
\State \textbf{return} $\omega_{\datasetX}$
\EndProcedure
\end{algorithmic}
\end{algorithm}

The next phase, shown in Algorithm~\ref{alg:fitbase}, is estimation of
the underlying base\_process that generates auto-correlation in the
time-series of the errors. This is done by using the CDF $B$ of the
beta distribution whose parameters have been inferred in step 1. Then
we operate a grid search over the $p$ and $q$ parameters to select the
order of the model that minimize the BIC criterion. We save the
coefficients. Recalling that we want the marginal of $Z \sim
\mathcal{N}(0,1)$, we set the variance of the errors of the base
process so that $Var[\mathbf{Z}_t] = 1$.  This procedure is explained
in section~\ref{sec:Correlation}.

\begin{algorithm}
\caption{Fitting the Base Process ARMA process \label{alg:fitbase}}
\begin{algorithmic}[1]
\Input $x$, $\ERR$, $\hat{\mathcal{S}}_x$
\Output $(a_i)_{i\leq p}, \: (b_i)_{i\leq q}, \: \sigma_{\delta}$
\Procedure{Fit\_Arma\_Process}{$x$, $\ERR$, $\hat{\mathcal{S}}_x$}
\For{$i \in [1, len(x)]$}\Comment{Estimating the base process see section~\ref{sec:Correlation}}
\State $ \hat{Z}_i\gets \phi^{-1}(B(\varepsilon_i, \mathcal{S}_{\mathcal{X}}(x_i)))$
\EndFor
\State $BIC\gets +\infty$
\State $p, \: q  \gets 0, \: 0$
\For{$p', q' \in [0, 5]^2$}\Comment{Grid Searching}
\State $tempBIC\gets BIC(ARMA(\hat{Z}, (p',0,q'))$
\If{$tempBIC<BIC$}
\State $BIC \gets tempBIC$
\State $p, \: q \gets p', \: q'$
\EndIf
\EndFor
\State $(a_i)_{i\leq p}, \: (b_i)_{i\leq q} \gets ARMA(\hat{Z}, (p,0,q))$
\State $\sigma_{\delta}\gets \argmin_{\sigma} (std(ARMA((a_i)_{i\leq p}, \: (b_i)_{i\leq q}, \sigma)-1)^2$
\State \textbf{return} $(a_i)_{i\leq p}, \: (b_i)_{i\leq p}, \: \sigma_{\delta}$
\EndProcedure
\end{algorithmic}
\end{algorithm}

\subsection{Procedures to deliver the target mare}

First, as shown in Algorithm~\ref{alg:infertarget}, given a target mare $\tilde{r}$, and a $\SID$ we verify that
$\tilde{r}$ is feasible. If it is, we aim at targeting a mean absolute
error for each conditional distribution with input in the $\SID$. For
this we compute a target function using the estimated weight function
(see section~\ref{plausibity_score}).

\begin{algorithm}
\caption{Inferring a target function for the SID \label{alg:infertarget}}
\begin{algorithmic}[1]
\Input $\hat{\mathcal{S}}$, $\SID, \tilde{r}, \hat{\omega}_{\datasetX}$
\Output $\tilde{m}$
\Procedure{Computing\_Simulation\_Target\_Function}{$\hat{\mathcal{S}}$, $\SID, \tilde{r}, \hat{\omega}_{\datasetX}$}
\State $ P_{\SID} \gets 0$\Comment{Computing the Plausibility score}
\For{$s \in \SID$}
\State $ P_{\SID} \gets  P_{\SID} + \frac{\hat{\omega}_{\datasetX}(s)}{|\SID|}$ .
\EndFor
\State $\tilde{r}_{max} \gets P_{\SID} \min\big(\frac{m_{max}(s)}{s\hat{\omega}_{\mathcal{X}}(s)}, s \in \SID \big) $
\If{$\tilde{r} > \tilde{r}_{max}$} \State \textbf{Report Error}
\EndIf
\State $\tilde{\omega}_{\SID}\gets \frac{\hat{\omega}_{\datasetX}}{P_{\SID}}$ 
\For{$s \in \SID$}\Comment{Applying the function as explained in section~\ref{plausibity_score}}
\State $\tilde{m}(s) \gets \tilde{r}s\tilde{\omega}_{\SID}(s)$
\EndFor
\State \textbf{return} $\tilde{m}$
\EndProcedure
\end{algorithmic}
\end{algorithm}

Second, as shown in Algorithm~\ref{alg:inferbeta}, according to a
target function $\tilde{m}$, we assign adjusted parameters for each
conditional distribution whose input is in the $\SID$. We move the
location parameters from the estimated ones while keeping the shape
parameters. See section~\ref{sec:adjust}.

\begin{algorithm}
\caption{Inferring the simulation beta distributions \label{alg:inferbeta}}
\begin{algorithmic}[1]
\Input $\tilde{m}$, $\mathcal{S}$, $\SID$
\Output $\tilde{\mathcal{S}}_{\tilde{m}}$
\Procedure{Adjusting\_Simulation\_Parameters}{$\tilde{m}$, $\mathcal{S}$, $\SID$}
\For{$\delta \in \SID$}\Comment{Applying the methodology explained in section}
\State $x \gets closest(\delta, \datasetX)$\Comment{not necessarily $\SID \subset \datasetX$}
\State $\tilde{\alpha}(\delta), \tilde{\beta}(\delta) \gets \hat{\alpha}(x), \hat{\beta}(x)$
\State $\tilde{l}(\delta), \tilde{s}(\delta) \gets Program_{1}(\tilde{\alpha}(\delta), \tilde{\beta}(\delta), \delta, \tilde{m}(\delta))$ \Comment{See equation~\ref{eq:w_prog}}
\State $\tilde{\mathcal{S}}_{\delta, \tilde{m}} = (\tilde{\alpha}(\delta), \tilde{\beta}(\delta), \tilde{l}(\delta), \tilde{s}(\delta))$
\EndFor
\State \textbf{return} $\tilde{\mathcal{S}}_{\tilde{m}}$ 
\EndProcedure
\end{algorithmic}
\end{algorithm}

\subsection{Procedure to simulate the output}

Using methods summarized in Algorithm~\ref{alg:simulate}, we simulate a base process sample of length $|\SID|$ and use
the simulated conditional distributions to obtain conditioned
errors. We directly get the simulation by summing the errors and the
input data. Finally, if the user asks for it, we optimize the curvature
{\em a posteriori}, see section~\ref{sec:Curvature}.

\begin{algorithm}
\caption{Simulating a sample of output \label{alg:simulate}}
\begin{algorithmic}[1]
\Input $\tilde{m}$, $\mathcal{S}$, $\SID, (a_i)_{i\leq p}, \: (b_i)_{i\leq p}, \: \sigma_{\delta}$, $ \tilde{\mathcal{S}}_{\SID}$, which implies $\tilde{F}_{\ERR|\X=x_t}^{-1}$
\Output $(\tilde{y}_i)_{i\leq n_{SID}}$
\Procedure{Computing\_Estimation\_Parameters}{$\tilde{m}, \mathcal{S}, \SID, (a_i)_{i\leq p}, \: (b_i)_{i\leq p}, \: \sigma_{\delta}$}
\State $(\tilde{z}_i)_{i\leq n_{SID}} \gets createArmaSample((a_i)_{i\leq p}, \: (b_i)_{i\leq p}, \: \sigma_{\delta}, \: n_{SID})$ 
\For{$i \in [1,n_{SID}]$}
\State $\tilde{\varepsilon}_i =\tilde{F}_{\ERR|\X=x_t}^{-1}(\phi(z_i))$
\State $\tilde{y}_i = x_i + \tilde{\varepsilon}_i$
\EndFor
\If{Curvature is $True$}
\State $(\tilde{y}_i)_{i\leq n_{SID}} \gets Optimization_1(\tilde{\varepsilon}, d, x, cap)$ \Comment{See Program~\ref{eq:prog_curvature}}
\EndIf
\State \textbf{return} $(\tilde{y}_i)_{i\leq n_{SID}}$ 
\EndProcedure
\end{algorithmic}
\end{algorithm}

\section{Evaluation}

We used computed scores to evaluate our simulations based on the
similarity with the empirical data and with the satisfaction of the
target. We want to assess the quality of the convergence of the
metrics with respect to : the number of scenarios simulated (let us denote it
$M$), the length of the input array (let us denote it $n_t$), the type
of simulation.

We study three types of simulations:

\begin{itemize}
	\item A) IID base process, $\phi_1$
	\item B) ARMA base process,  $\phi_2$
	\item C) ARMA base process and curvature optimization,  $\phi_3$
\end{itemize}

\subsection{Target MAPE Score}

The score function for achieving the target MAPE is
$$S_{mare}(M,n_t,k) = \sqrt{\sum_{i=1}^{M}( \tilde{r}*100\% - MAPE((x_i)_{i \leq n_t}, \phi_k((x_i)_{i \leq n_t})))^2}.$$

\subsection{Validation of the base process}
Let $p$ be the maximum lag of auto-correlation we wish to assess.

Let us define the functional $$\hat{\rho}((\varepsilon_i)_{i \leq n}, j) =  \frac{1}{(n-j)\sigma^2}\sum_{i=0}^{n-j}\varepsilon_{i+j}\varepsilon_{i}$$

If $\bar{\varepsilon} = 0$, $\hat{\rho}((\varepsilon_i)_{i \leq n}, j)$ is the estimation of the auto-correlation of the errors of the input dataset at lag $j$. $\hat{\rho}((\phi_k(x_i)-x_i)_{i \leq n_t} , j)$ is the estimation of the auto-correlation at lag $j$ of the errors simulated by the MAPE\_maker of type $k$ with an SID starting at the beginning of the dataset and of length $n_t$.

$$S_{auto\_correlation}(M,n_t,k,p) = \sqrt{\sum_{i=1}^{M} \sum_{j=1}^{p} ( \hat{\rho}((\varepsilon_i)_{i \leq n}, j) - \hat{\rho}((\phi_k(x_i)-x_i)_{i \leq n_t}, j))^2}$$

\subsection{Validation of the curvature}

Let us define the functional $$D((y_i)_{i \leq n}) =  \frac{1}{n-2}\sum_{i=0}^{n-2} y_i-2y_{i-1} + y_{i-2}$$

$$S_{second\_difference}(M,n_t,k) = \sqrt{\sum_{i=1}^{M} (D((y_i)_{i \leq n}) - D(\phi_k(x_i)_{i \leq n_t}))^2}$$

\subsection{Score function}

The score function is the sum of those three targets weighted :

\begin{align*}
S(M,n_t,k,p; \: w_m,  w_{ac},  w_{sd}) &=  w_m*S_{mare}(M,n_t,k) + \\ 
																				&		 w_{ac}*S_{auto\_correlation}(M,n_t,k,p) + \\
																				&		 w_{sd}*S_{second\_difference}(M,n_t,k) 
\end{align*}

\subsection{Behavior of the scores as $M$ grows}

\begin{figure}[h]
    \centering
    \includegraphics[scale=0.75]{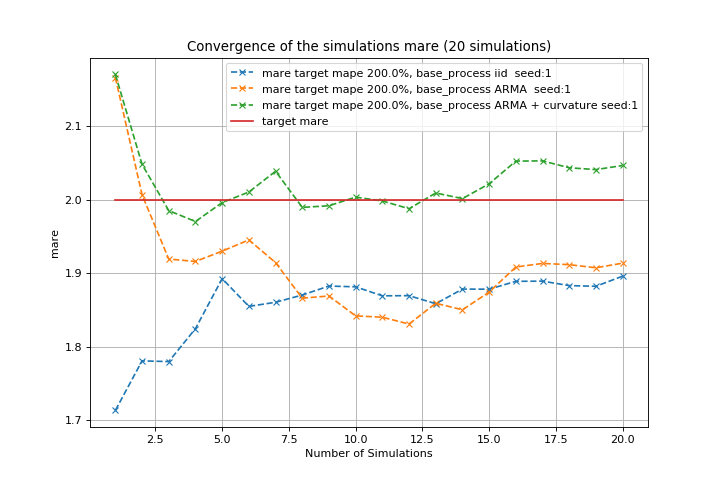}
    \caption{MARE score as a function of the number of scenarios created by simulation.}
    \label{fig:mapeconverge}
\end{figure}

\begin{figure}[h]
    \centering
    \includegraphics[scale=0.75]{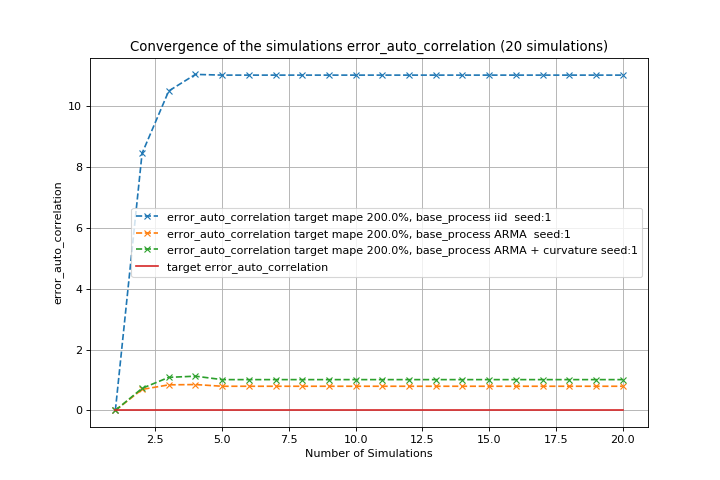}
    \caption{Sum of the absolute difference of the auto-correlation score as a function of the number of scenarios created by simulation.}
    \label{fig:autoconverge}
\end{figure}

To illustrate the behavior of the simulated scenarios as the number of
scenarios created, $M$, grows we conducted experiments using the CAISO
wind dataset and created scenarios for three
days. Figure~\ref{fig:mapeconverge} shows that for this example, the
achieved MARE is close to the target MARE as soon as there are about 4
scenarios. The scenarios that use curvature correction result in a
value that is closest to the target, which makes sense because
Program~\ref{eq:prog_curvature} corrects for the MARE after the
scenarios are created. However, the other scenarios are reasonably
close. Figure~\ref{fig:autoconverge} demonstrates that for processes
that are not iid, the autocorrelation score is quit good almost
regardless of the number of scenarios. Figure~/ref{fig:curveconverge}
shows that the curvature (second differences) score does not depend on
the number of scenarios generated and that the methods are ordered as
expected.

\begin{figure}[h]
    \centering
    \includegraphics[scale=0.75]{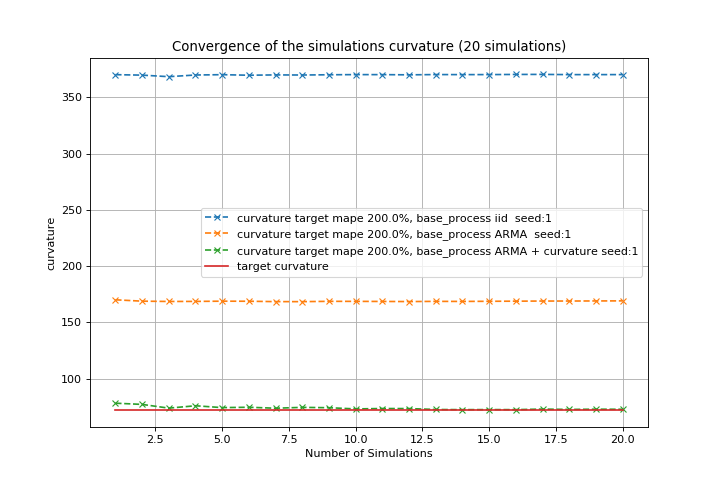}
    \caption{Second difference score as a function of the number of scenarios created by simulation.}
    \label{fig:curveconverge}
\end{figure}

\section{Conclusions}

We have described methods for creating scenarios that make use of a
history of forecast errors. The corresponding software is available
for download and use.  Although we used wind data from CAISO in our
illustrations, the method can be used for any situation where there is
a history of forecasts and actuals. In particular, the software has
been used to create scenarios for load, solar, and wind for the
rts-gmlc data \url{https://github.com/GridMod/RTS-GMLC}.

The use of solar requires pre- and post-processing of the input data to work well. Instead
of power values, the forecasts and actuals should be presented as fractions of capacity
and with the value of $cap$ set to one during the day and zero at night. This is because
solar power is always zero at night and because the concept of ``low power'' changes
during the day.

Future research includes consideration of error measures other than
the MAPE. On the purely software front, we are working to parallelize
computations. The software and the methods described here are intended
to be an addition to the kit of tools available for dealing with
uncertainty in power generation planning and operations.

\section*{Acknowledgement}

Sandia National Laboratories is a multi-mission laboratory managed and
operated by National Technology and Engineering Solutions of Sandia,
LLC., a wholly owned subsidiary of Honeywell International, Inc., for
the U.S. Department of Energy’s National Nuclear Security
Administration under contract DE-NA0003525. This paper describes
objective technical results and analysis. Any subjective views or
opinions that might be expressed in the paper do not necessarily
represent the views of the U.S. Department of Energy or the United
States Government.

\bibliographystyle{plain}
\bibliography{refs}

\end{document}